\documentstyle[aps,epsf,epsfig]{revtex}
\begin{document}
%\draft
\twocolumn[\hsize\textwidth\columnwidth\hsize\csname @twocolumnfalse\endcsname 
\title{Constraints from $T_c$ and the isotope effect for MgB$_2$}
\author{A. Knigavko and F. Marsiglio}

\address{Department of Physics, University of Alberta, Edmonton,
Alberta, Canada T6G 2J1}

\date{\today} 
\maketitle 
\begin{abstract} 
With the constraint that $T_c = 39$ \, K, as observed for MgB$_2$, we use
the Eliashberg equations to compute possible allowed values of the isotope
coefficient, $\beta$. We find that while the observed value $\beta_{\rm obs}
= 0.32 \pm 0.01$ can be obtained in principle, it is difficult to reconcile
a recently calculated spectral function with such a low observed value.
\end{abstract}

\vskip2pc]
 
\section{Introduction}
The discovery of an isotope effect in superconducting Hg 
\cite{maxwell50,reynolds50} had a significant influence on the

subsequent development of the theory of superconductivity. Since that
time the isotope effect has been measured for almost every superconductor
where possible, with results that span the range of complete agreement
with BCS expectations to a null result \cite{meservey69}. In this 
latter instance, the implications for mechanism have differed, depending

on the circumstance. For example if $T_c$ is low, arguments based on
competition of the electron-phonon induced attractive coupling with 
the repulsive Coulomb pseudopotential suggest that the isotope
coefficient should be close to zero in this case \cite{morel62,garland63}.

Qualitatively, the argument is as follows: increasing the phonon
frequency (by reducing the ionic mass) serves to increase $T_c$ because
the energy scale that governs $T_c$ is the phonon energy. On the other
hand $T_c$ is also largely determined by the ability of phonons to
mediate a {\it retarded} attractive interaction that avoids the
{\it instantaneous} Coulomb repulsive potential. This requires a
large difference between the phonon energy scale (approximately the
Debye frequency, $\omega_D$) and the energy scale of the Coulomb
potential (roughly half the bandwidth, $D/2$). Raising the phonon energy
scale (through a lower mass isotopic substitution) brings these two
energy scales closer together, and therefore acts to {\it reduce} $T_c$.
This effect offsets the gain in $T_c$ expected due to the first argument,
and can lead, in some cases to an isotope coefficient close to zero.

The discovery of superconductivity in MgB$_2$ at 39 K \cite{nagamatsu01} has naturally prompted 
questions about mechanism. Isotopic substitution of both Mg and B has
resulted in an overall isotope coefficient of $0.32 \pm 0.01$ 
\cite{budko01,hinks01}. Assuming the principle of superposition holds for
the isotopic substitutions, in the absence of Coulomb repulsion,
a value of $\beta = 0.5$ is expected. Here,
$\beta \equiv \sum_i \beta_i$, where $\beta_i \equiv -d \ln T_c/d \ln M_i$, 
and $M_i$ is the mass of the i$^{th}$ ion, and the summation is over all the 
ions in the compound \cite{rainer79}. The simplest version of
Eliashberg theory \cite{carbotte90} then requires a Coulomb pseudopotential
to explain this reduced value. 

In fact, as explained there, and more recently in Ref. \cite{hinks01},
one can gain a qualitative understanding of this reduction through the
modified McMillan equation \cite{mcmillan68,allen75},
\begin{equation}
k_B T_c = {\hbar \omega_{\ln} \over 1.2} \exp{\biggl(- {1.04(1 + \lambda)
\over \lambda - \mu^\ast (1 + 0.62 \lambda) } \biggr)}.
\label{mcmill}
\end{equation}        
Here $\lambda$ is the electron-phonon coupling constant, $\omega_{\ln}$
is the logarithmically weighted average phonon frequency, and $\mu^\ast$
is the Coulomb pseudopotential. It is important to note, for comparisons
which we will make later, that the McMillan equation was derived with
$\mu^\ast \equiv \mu^\ast(\omega_{\ln})$. That is, the cutoff associated
with the pseudopotential is the characteristic phonon frequency itself.

The approximate formula for obtaining the pseudopotential at any frequency
$\omega_i$, given the `bare' potential $\mu$ associated with the bandwidth
energy scale, $D/2$, is
\begin{equation}
\mu^\ast(\omega_i) \approx {\mu(D/2)\over 1+\mu(D/2) \ln{D/2 \over \omega_i} }.
\label{pseudo}
\end{equation}
Equation \ref{pseudo} requires assumptions of particle-hole symmetry and the
smallness of the parameter $T_c/\omega_i$. Removal of these assumptions is
further discussed in Ref. \cite{marsiglio89}. Most importantly, however,
for this relation to work requires $\omega_i {> \atop \sim} 6 \omega_{\ln}$.
Nonetheless, for purposes of comparing numerical solutions of Eliashberg
theory to the McMillan equation, we will apply Eq. \ref{pseudo} for
$\omega_i = \omega_{\ln}$.

Based on Eqs. (\ref{mcmill}) and (\ref{pseudo}) the isotope coefficient
is given by

\begin{equation}
\beta = {1 \over 2} \biggl(
1 - {1.04(1 + \lambda)(1 + 0.62\lambda) \over [\lambda - \mu^\ast (1
+ 0.62 \lambda)]^2} \mu^{\ast 2}
\biggr).
\label{mcmill_beta}
\end{equation}            
These equations have been used previously to constrain $\lambda$ and $\mu^\ast$
\cite{carbotte90}. Very recently Hinks et al. \cite{hinks01} have
argued that their observed isotope effect is incompatible with a small value
of $\mu^\ast$, as suggested by calculations of the electron-phonon parameters \cite{remark0}.
The purpose of this paper is to examine these conclusions with numerical
solutions to the Eliashberg equations. In this way not only can one examine
the accuracy of the McMillan equation, but also the dependency on details 
of the electron-phonon spectral function, $\alpha^2F(\omega)$ (recall 
that the fitted parameters in the McMillan equation were initially based on the
Nb spectrum). In the following section we will first examine solutions
for an Einstein function spectrum, for which parameters can be varied simply
and systematically. We find, upon comparison with the results from the
McMillan equation, reasonably good agreement. In Section III we examine
first the effect of broadening this spectrum, and then the effect
of using a more realistic spectrum, namely the one which has been calculated
with {\it ab initio} methods \cite{kong01}. A summary is provided in the
final section.

\section{Einstein spectrum}

Inspection of the calculated spectrum in Ref. \cite{kong01} shows a
spectrum that is dominated by one peak at approximately 75 meV. While we
will later explore the consequences of this spectrum, for now we simply
model $\alpha^2F(\omega)$ with an Einstein spectrum:
\begin{equation}
\alpha^2F(\omega) = {\lambda \Omega \over 2} \delta(\omega - \Omega).
\label{ein}
\end{equation}  
Based on the known values of $T_c$ (39 K) and $\beta$ (0.32), then, for a given
phonon frequency and value of $\mu$, one can determine the electron-phonon coupling
strength, characterized by $\lambda$. 

Because we will vary the phonon spectrum characteristics
in what follows (in particular the frequency scale), we choose to characterize
the direct Coulomb potential by its value at some electronic energy scale
associated with the bandwidth \cite{remark1}; we choose this energy scale to be
$D/2 = 5$\, eV, to correspond roughly with the bandwidth of MgB$_2$ \cite{kortus01,an01}.
In Fig. 1 we show the results of this exercise. In Fig. 1a we first show
$\mu^\ast(\Omega)$ vs. $\mu(D/2 = 5 {\rm eV})$, for a variety of Einstein frequencies.
These curves are calculated directly from Eq. (\ref{pseudo}); note, however, that

a different cutoff frequency has been used for each curve, corresponding to the
phonon frequency for that particular case. It means that as $\mu(D/2)$ increases (in
principle without bound), $\mu^\ast(\Omega)$ saturates, and in fact, can barely exceed a
value $\approx 0.2$ \cite{remark2}. Note that for $\mu \approx 2$, then, with 
$D \approx 10$\, eV,  the effective Hubbard $U$ is already approximately 20 eV.

In Fig. 1b the values of $\lambda$ required to yield $T_c = 39$\, K are shown as 
a function of $\mu(D/2)$. Note that as $\mu$ (henceforth we will denote $\mu(D/2)$
by simply $\mu$) increases $\lambda$ also saturates, specifically because, as
shown in Fig. 1a, $\mu^\ast(\Omega)$ saturates. Moreover, the saturated values
vary with frequency, but for the most part are reasonably moderate, of order unity
\cite{alexandrov01}.

Finally, in Fig. 1c we show the calculated isotope coefficient, $\beta$, as a 
function of $\mu$. These of course begin at 0.5 (for $\mu = 0$), and fall gradually,
while also showing some signs of saturation (again, as a consequence of Fig. 1a).
This figure restricts the possible parameters; we obtain an isotope coefficient 
close to the observed result with a sufficiently high phonon frequency.

Thus, the observed isotope shift in MgB$_2$ can readily be obtained within the
framework of Eliashberg theory. Visual inspection of Fig. 1 shows that the value
of $\mu^\ast(\Omega)$ must be `high', i.e.  it must exceed the `canonical' value
of 0.1 (as is often adopted in studies of $T_c$). On the other hand, the
value required is not excessive, i.e. $\mu^\ast(\Omega) \approx 0.2$ is sufficient.
Moreover, the required value of $\lambda$ is moderate, as already commented on. The
last restriction is the phonon frequency itself. Here, we require fairly high
phonon frequencies, with $\Omega \approx 100$\, meV. This latter constraint poses
two difficulties. First, the required value must exceed that calculated \cite{kong01}
by a small amount (in addition the coupling strength here exceeds that calculated by a
significant amount as well). Secondly, even if we disregard the {\it ab initio} calculation
the possibility of high frequency phonons coupling to electrons with $\lambda \approx 1$
has been questioned in the past \cite{cohen72}. We are unable to resolve the issue of the
last comment; however, before we suggest that a discrepancy exist with the {\it ab initio}
work, we should further investigate the impact of spectral function shape on our
results. We address this in the next section.

Before we leave this section, however, it is of interest to compare our results with 
those obtained by the McMillan equations (\ref{mcmill}) and (\ref{mcmill_beta}), as
used in Ref. \cite{hinks01}. In Fig. 2 we plot (a) $\lambda$, and (b) $\beta$ vs.
$\mu$, for a few select Einstein frequencies. For the McMillan equations, we have 
used Eq. (\ref{pseudo}) (with $\omega_i = \Omega$) to determine $\mu^\ast(\Omega)$. 
It is clear that the McMillan equation is excellent in this regime; the value
of $\lambda$ is accurate to within 5\%, while the value of $\beta$ is somewhat
over-estimated by the McMillan equation, but follows the trends given by the
numerical solutions to the Eliashberg equations fairly closely.

Nonetheless, because it is important to explore spectral
shape dependence, we continue with our use of the numerical Eliashberg equations
\cite{carbotte90}.

\section{Lorentzian spectrum, and ab initio spectrum}

We begin first with some systematics. To understand how a realistic spectrum
impacts our conclusions, we adopt the following straightforward procedure.
First, we continuously broaden the Einstein mode at a given frequency into a
Lorentzian with mean frequency $\Omega$ and half-width $\sigma$ (see Fig. 3a). 
We begin at a particular point on the curves in Fig. 1. For example, if we take 
$\mu = 1.2$ and $\Omega = 75$\, meV, then $\lambda = 1.08$ and $\beta = 0.355$. 
Now we imagine that $\mu$ is fixed, and upon broadening, the value of $\lambda$ is
readjusted (in this case it is increased) to maintain $T_c = 39$\, K.
We now calculate $\beta$ (as before, except now we are required to shift an entire 
spectrum) to see if a broadening of the spectral function decreases or increases
$\beta$. 

The solid curve in Fig. 3c shows our results for the parameter regime used above 
($\mu = 1.2$ and $\Omega = 75$\, meV). For calculations the values of sigma were 
chosen to be 0.1, 2, 6, and 40. Clearly the effect of spectral broadening is small, 
and, in this regime, works to raise
the isotope coefficient. We thus anticipate that adopting a more realistic spectrum
will take us further from the experimental value. Since the mean frequency is
rather high, we also broaden the spectrum asymmetrically, as indicated in Fig. 3b.
The result is that $\beta$ is increased even further, as shown by the dashed line 
in Fig. 3c. While we have no proof, we suspect that for a given $T_c$ and $\mu$, an 
Einstein spectrum represents a lower bound for the value of $\beta$. This allows us 
to use Fig. 1 to make definitive statements concerning the possible value of $\beta$,
obtained for a given $T_c$ and $\mu$ (or, alternatively, $\lambda$).

We take as a most plausible spectrum for MgB$_2$ the one calculated in Ref. \cite{kong01}.
Using the absolute value of the coupling $\lambda=0.836$ calculated from this spectrum 
will lead to an isotope coefficient very close to 0.5, in clear disagreement 
with experiment \cite{remark3}.
In fact this can be easily estimated on the basis of Fig. 1. With a value of
$\mu^\ast(\omega_c = 1 eV) \approx 0.1$, then $\mu \approx 0.12$, and so Fig. 1
gives $\beta {> \atop \sim} 0.46$, regardless of what the characteristic frequency
of the spectrum is.

To reconcile this discrepancy, we adjust the overall coupling strength (maintaining
their calculated spectral shape) in an attempt to achieve $\beta \approx 0.32$
with minimal change. As Fig. 1 shows, we will need to increase $\mu$, and hence
$\lambda$. Thus, we assume that the calculated spectral strength has somehow been
underestimated. The calculated dependence of $\beta$ on $\mu$ is shown by
the dashed 
line in Fig. 4. 
We find that $\beta$ remains always above 0.355 for $\mu < 2$ (i.e.
$U {< \atop \sim} 20$ \, eV), and, for more moderate values of $\mu$, is
greater than 0.4.

\section{Summary}

Fig. 1 illustrates that in principle, an isotope coefficient $\beta \approx 0.32$
with $T_c = 39$ \, K is possible with a modest value of $\lambda$. However, this does not
take into account the objection raised long ago \cite{cohen72} to the possibility of
a system to maintain high phonon frequencies and moderately strong electron-phonon
coupling. The use of a more realistic spectrum reinforces this difficulty. The
value of $\beta$ is raised, and so a spectrum with an even higher characteristic
phonon frequency is required. Finally, using the spectral shape calculated in 
Ref. \cite{kong01} yields an even higher value of $\beta$, so that a value of
$\beta \approx 0.4$ is obtained, in clear {\it disagreement} with experiment. Possible
solutions to this discrepancy might include the effect of the band edge about 
0.5 eV from the Fermi energy. As discussed in the second entry of Ref. \cite{marsiglio89}
a low hole concentration leads to a reduced isotope effect. However, for MgB$_2$ the
energy difference between the band edge and the Fermi level is still about 5-10 times
the characteristic phonon frequency. A second possibility is the role of
anharmonicity \cite{yildirim01} in the primary phonon which couples to the
electron. Unfortunately, no formalism presently exists that allows us to
compute $T_c$ from an electron-phonon interaction in the case where the phonons are
anharmonic. Finally, there remains the possibility that the mechanism is not the
electron-phonon interaction \cite{hirsch01}. Some enlightening data would be a measurement
of the isotope coefficient for either hole or electron doped MgB$_2$. The primary effect
would be  $T_c(n)$ itself, as has been discussed in Refs. \cite{hirsch01,hirsch01a}, 
but a measurement of $\beta(n)$ would also help differentiate between doping effects
due to the electron-phonon interaction, and doping effects due to some other
mechanism.

\acknowledgements
This work was supported by the Natural Sciences and Engineering
Research Council (NSERC) of Canada and the Canadian Institute for Advanced
Research.

\onecolumn

\begin{figure}[h]
\epsfig{figure=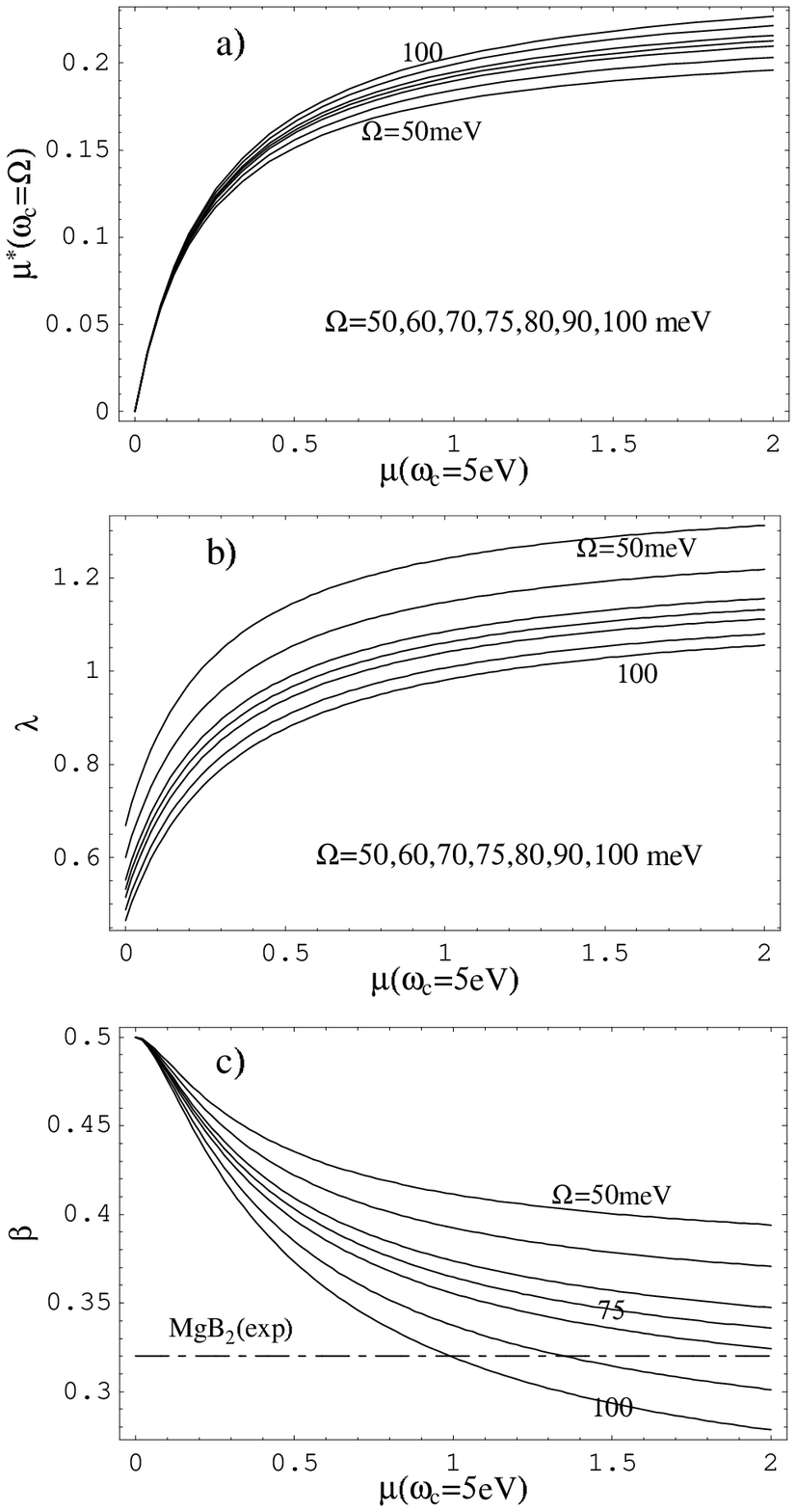,height=20.5cm,width=12.5cm}
\caption {a) The pseudopotential $\mu^\ast(\omega_c = \Omega)$, 
b) electron--phonon coupling constant $\lambda$, and 
c) isotope coefficient $\beta$, as a function of $\mu$. The value of $\mu$ 
is varied while $T_c$ is held fixed at 39 K. These curves represent solutions 
to the linearized Eliashberg equations with a cutoff $\omega_c = 5$\, eV, 
with an Einstein spectrum with a frequency as shown. Smaller values of 
$\beta$ are obtained only for spectra with high frequencies, and, 
in that case, with high values of $\mu$.
}
\label{Fig1}
\end{figure}

\begin{figure}
\epsfig{figure=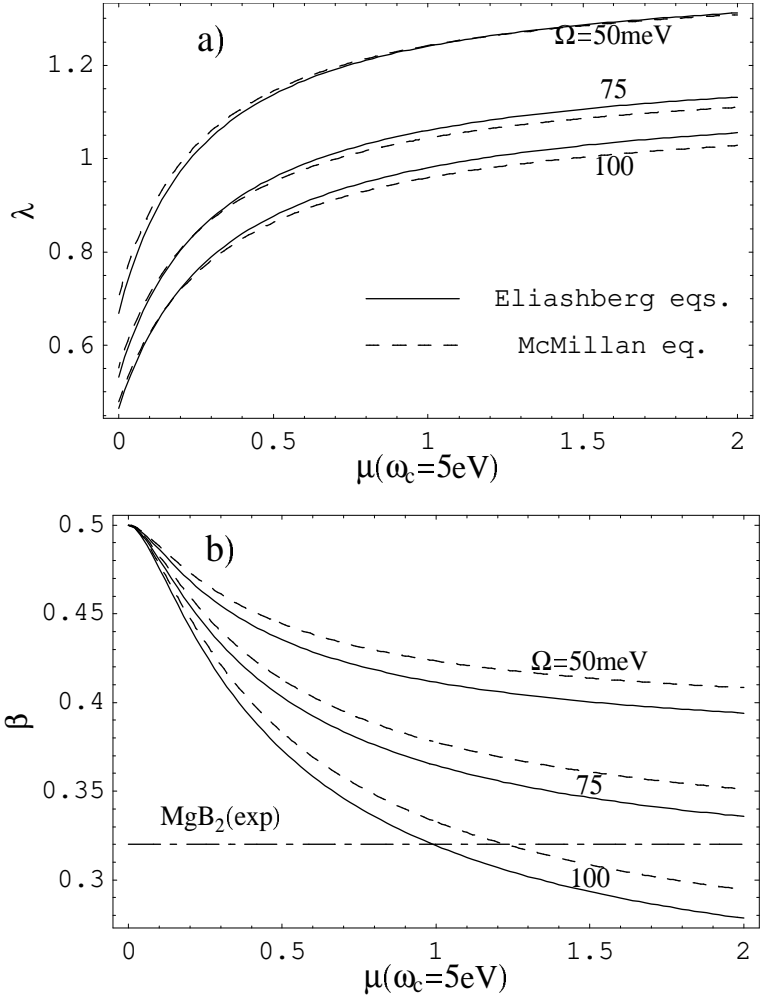,height=14.0cm,width=10.5cm}
\caption {Comparison of the value of a) $\lambda$, and b) $\beta$, obtained 
with the McMillan equation compared to a full Eliashberg solution for an Einstein
spectrum, as in Fig. 1. For all values of $\mu$ the agreement is excellent for $\lambda$,
and qualitatively very good for $\beta$.
}
\label{Fig2}
\end{figure}   

\begin{figure}
\epsfig{figure=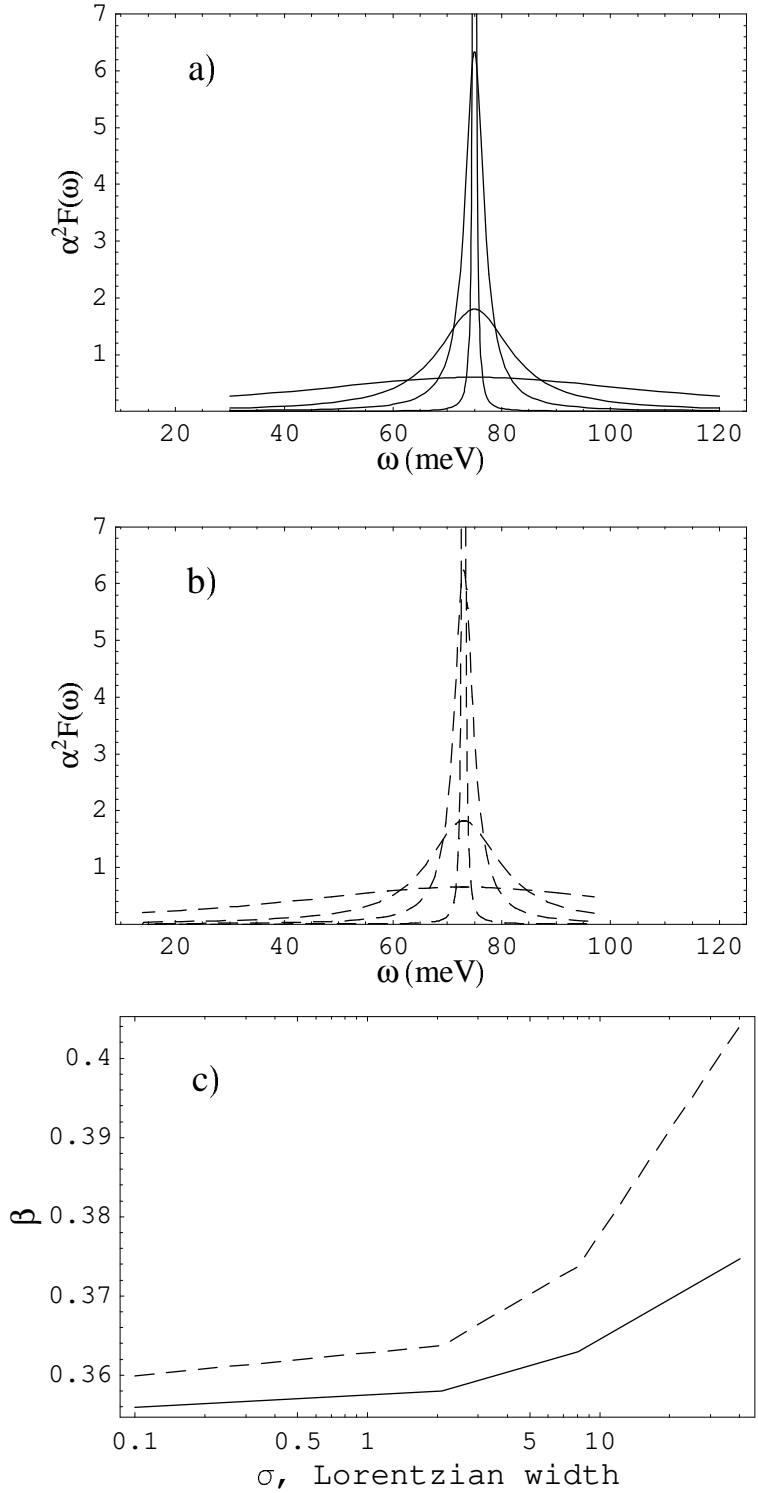,height=20.5cm,width=10.5cm}
\caption {Dependence of isotope coefficient $\beta$ on the broadening parameter $\sigma$. 
In a) and b) the actual spectra used are drawn (see text for detail). In c) the value 
of $\beta$ is plotted as a function of $\sigma$. Clearly, $\beta$ increases with broadening 
(while $T_c$ and $\mu$ are held fixed).
}
\label{Fig3}
\end{figure}

\begin{figure}
\epsfig{figure=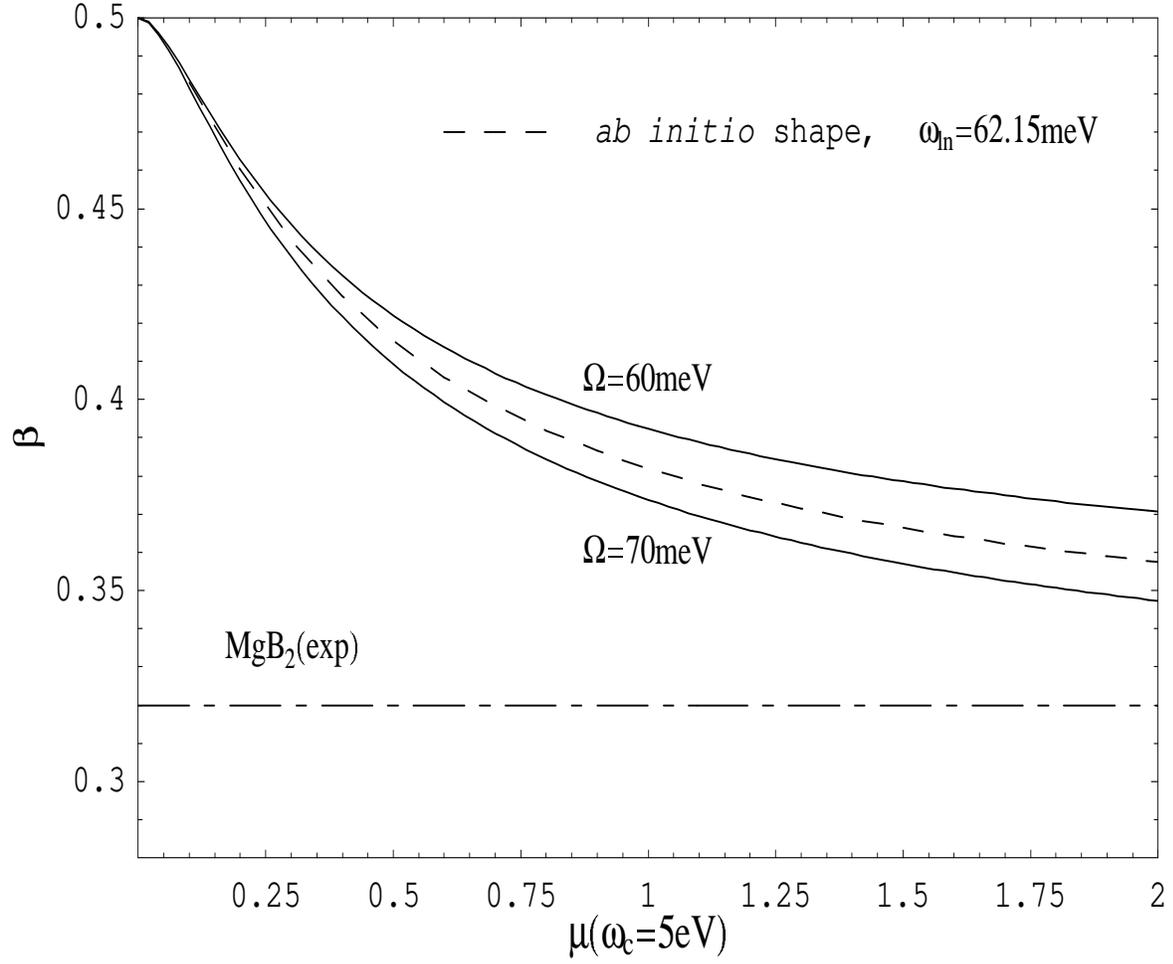,height=14.5cm,width=16.0cm}
\caption {$\beta$ as a function of $\mu$, as determined by scaling the calculated
spectrum of Kong {\it et al.} \protect\cite{kong01} to keep $T_c = 39$ \, K. To
achieve $\beta < 0.4$ requires a very large value of $\mu$. The experimentally
observed value of $\beta = 0.32 \pm 0.01$ is very difficult to obtain with this
calculated spectrum. Results for the Einstein spectrum with $\Omega =$ 60 and 70 meV are
also shown for comparison.
}
\label{Fig4}
\end{figure}


\begin{references}
\bibitem{remark0} This argument was particularly forceful when the calculated
value of $\lambda$ led to a very small value of $\mu^\ast = 0.02$. A more
accurate calculation leads to $\mu^\ast = 0.1$ \cite{kong01}.
\bibitem{remark1} Again we refer the reader to Ref. \cite{marsiglio89} for
a more careful treatment of the direct Coulomb repulsion, particularly
for the case when the chemical potential approaches a bandedge.
\bibitem{remark2} This value is determined by the limit in which $\mu \rightarrow \infty$, i.e.
$1/\ln{D/2/\Omega}$. For typical values of parameters this will not exceed the value 0.2 by
a significant amount.
\end{references}
\end{document}